\documentclass[11pt,twoside]{article}
\usepackage{asp2010,graphicx}

\resetcounters

\begin{document}

\title{The multiple periods and the magnetic nature of CP\,Pup}
\author{Elena Mason,$^1$, Antonio Bianchini$^2$, Marina Orio$^{3,7}$, Robert E. Williams$^1$, Koji Mukai$^4$,  Domitilla de Marino$^5$, Timothy .M.C. Abbot$^6$, Francesco di Mille$^2$
\affil{$^1$STScI, Baltimore, MD 21218, USA}
\affil{$^2$Universita` di Padova, vicolo  dell'Osservatorio 3, I-35122 Padova, IT}
\affil{$^3$INAF Padova, vicolo  dell'Osservatorio 5, I-35122 Padova, IT}
\affil{$^4$NASA Goddard Space Flight Center, Greenbelt, MD 20771, USA}
\affil{$^5$INAF Capodimonte, salita Moiariello 16, 80131, Napoli, IT}
\affil{$^6$CTIO, Casilla 603, La Serena, Chile}
\affil{$^7$University of Wisconsin, 475 N. Charter Str., Madison, WI 53704, USA}
}

\begin{abstract}
Fast  cadence  time  resolved  spectra taken at the CTIO-4\,m telescope with the RC-­spectrograph  during  2  consecutive  nights  revealed  a long  term  modulation  of  the  binary  radial  velocity. Chandra  hard  X-­ray  spectra  taken  with  the  HETGS 
instrument showed features typically observed in magnetic white dwarfs (WD). Here, we present the new data  and  suggest that CP\,Pup is possibly a long orbital period intermediate polar. 
\end{abstract}

\section{Introduction}

CP\,Pup  has  been one of  the fastest  and brightest  novae  ever  recorded. It reached  V$_{max}$=­0.2\,mag in Nov 1942 and declined by several mag in less than 1 week ($t_3$=6.5 days). After outburst it has not returned to the quiescent level it had before outburst, remaining about 5 mag brighter (Schaefer \& Collazzi 2010).

The  post  outburst  quiescent  phase  has  been  studied through  time  resolved  broad  band  imaging  and spectroscopy  (Barrera \& Vogt 1989, Bianchini et al. 1985, 2012, Cropper 1986, Diaz \& Steiner 1991, Duerbeck et al. 1987, O'Donogue et al. 1989, Patterson \& Warner 1998, Szkody \& Feinswog 1988, Warner 1985, White et al. 1993) suggesting that CP\,Pup is a short orbital period Cataclysmic Variable (CV).  The  spectroscopic  period  is shorter  than  the photometric period and both periods are unstable/variable.  The  WD  mass derived  from  radial velocity  studies  has  always  been  too  small  ($<$0.2 to $<$0.6\,M$_\odot$) to be consistent with classical nova theory and stellar  evolution  time  scales.  X-ray  observations (Balman et al. 1995, Orio et al. 2009) have always  suggested  a  possible  magnetic  nature  of  the primary WD,  though  they could never  firmly  establish  it.

\section{Optical observations and results}
Time  resolved  spectra  were collected at  the CTIO  4m telescope equipped with the 
RC-spectrograph on Feb 6 and 7 2009. The wavelength coverage was 3500-­6000\,\AA\, with a resolution FWHM$\sim$6\,\AA. The exposure times were 60\,s each with a duty cycle 
of less than 2\,min. 

The radial velocity measures of the H$\beta$, H$\gamma$, He\,{\sc ii}$\lambda$4686 and C\,{\sc iii}(1), show a drift of the data points during the first night and possibly an offset from night to night (Fig.\,1). From the Fourier analysis of the whole data set we find two periods (Fig.\,2): 1) the well known $\sim$1.47\,hr period (up to now interpreted as the orbital period of the binary), and 2) a longer period of $\sim$10\,hr. This period is highly uncertain but possibly matches the true orbital period of the system. 

\begin{figure}
\centering
\includegraphics[width=10cm,angle=270]{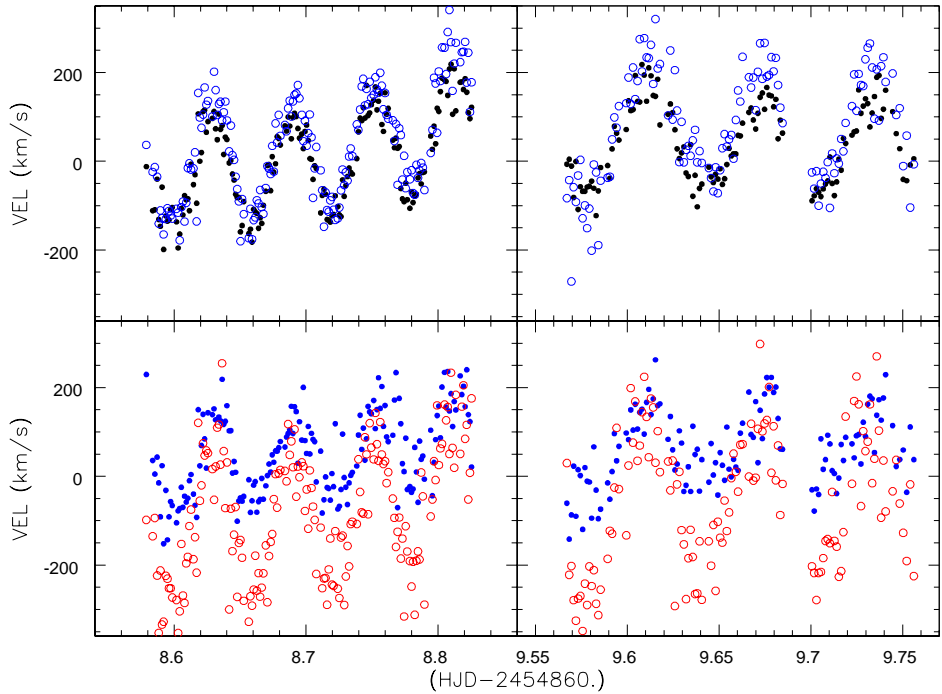}
\caption{CP\,Pup radial velocities observed during the nights Feb 6 (left panels) and Feb 7 (right panels), 2009. H$\beta$ (black symbols) and H$\gamma$ (blue circles) radial velocities are shown in the top panels; while the He\,{\sc ii} (blue symbols) and C\,{\sc iii} (red circles) lines are shown in the bottom panels.  }
\label{RV}%
\end{figure}

\begin{figure}
\centering
\includegraphics[width=5.truecm, angle=0]{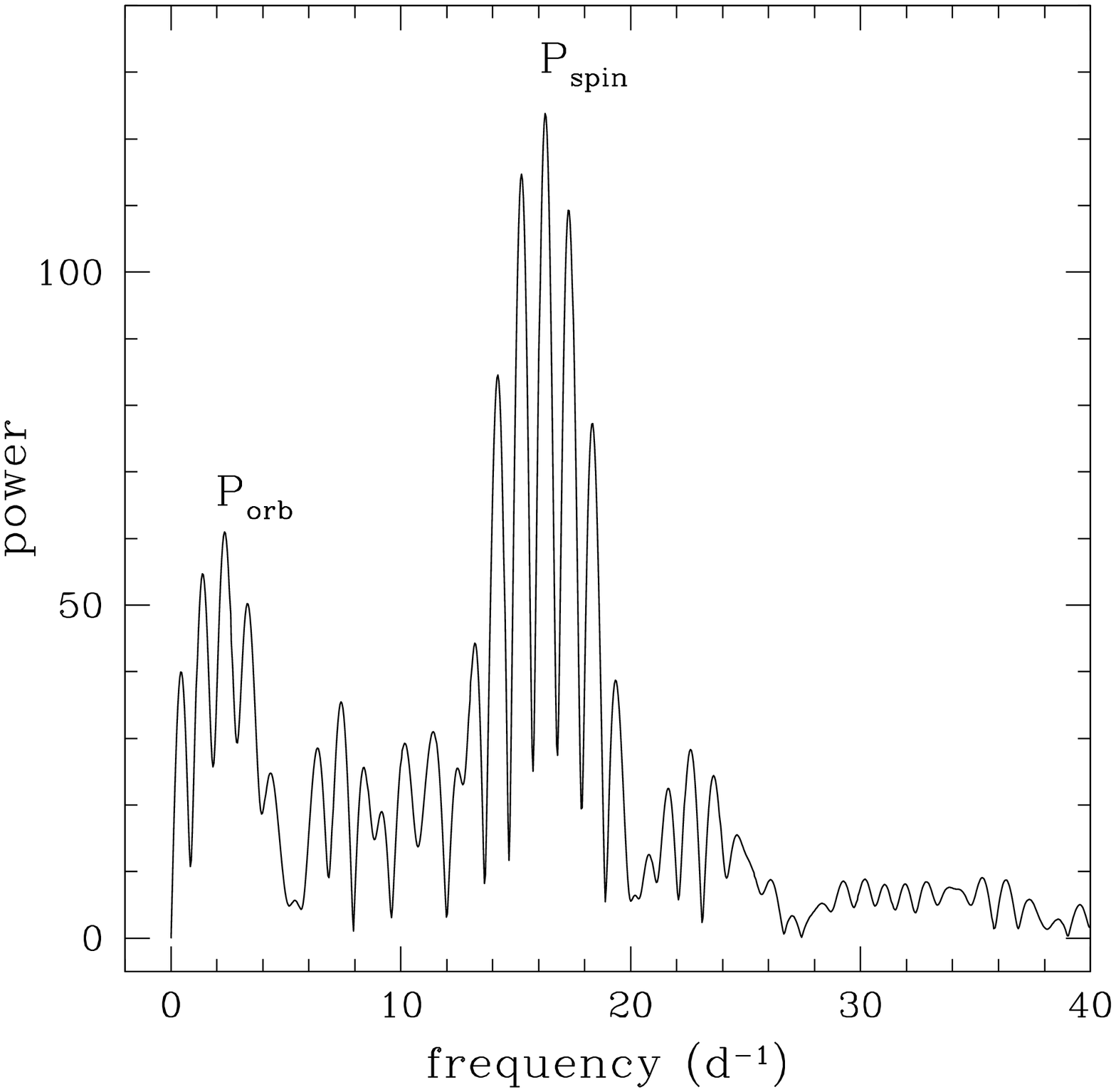}
\includegraphics[width=5.truecm, angle=0]{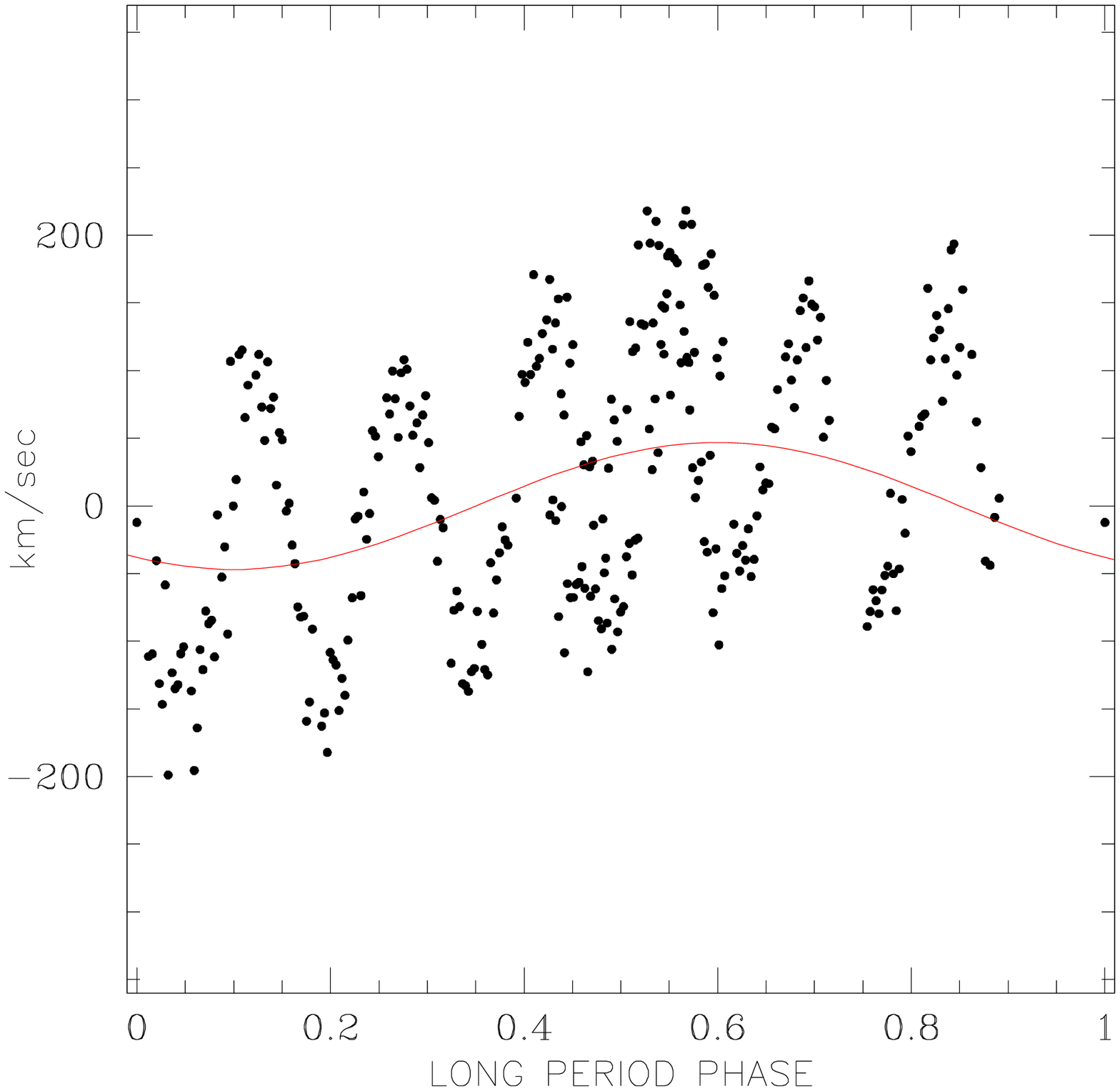}
\caption{Left panel: Power spectrum of the 2 nights H$\beta$ radial velocities. Right panel: radial velocity curve of the H$\beta$ emission line phased on the long  period of $\sim$9.77\,hr and best fit (red line).}
\label{FT}%
\end{figure}

\section{X-ray observations and results}

High resolution X-ray spectra were taken with Chandra and the HETGS spectrograph. The wavelength coverage was 0.4­10\,keV at a resolution of E/$\Delta$E$\sim$1000.  Seven  exposures  were 
collected  between  September 30 and October 31 2009 for a 175780\,s of total integration time. 

Our new Chandra observations -- though limited by a relatively low signal-to-noise ratio -- show that the X-ray spectrum of CP\,Pup is modified by a complex, intrinsic absorber.  While this is commonly seen in magnetic CVs, it has not been seen in non-magnetic
CVs except at very high inclination angle (i.e., systems showing at least a grazing eclipse; see Mukai et al. 2009). This supports the magnetic CV interpretation. Within this hypothesis and following Mukai et al. (2003) we fit the spectra using a cooling 
flow model. Our best fit model temperature is kT=36.5\,keV with a 90\% confidence range of 20.2-­55.7\,keV.  Our best fit temperature corresponds to a WD mass of 0.8 (0.57-­0.99) M$_\odot$. Such a WD mass is in much better agreement with the current classical nova theory, though still on the low end of its expected value.

\begin{figure}
\centering
\includegraphics[width=10cm,angle=0]{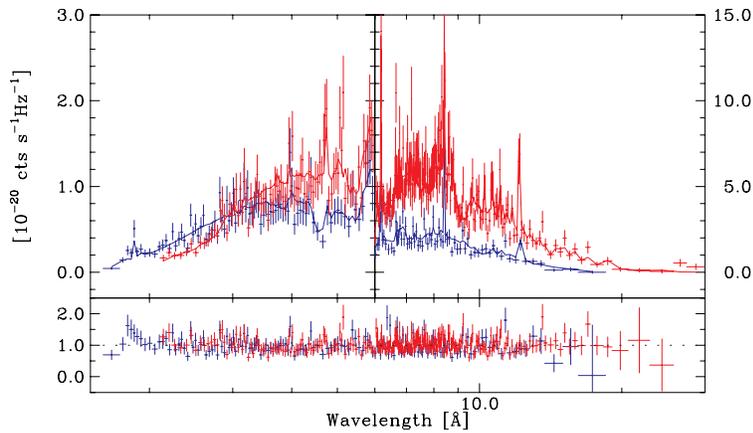}
\caption{Upper panels: data and model in logarithmic (blue) wavelength and linear (red)
count rate scale.  Note that we split the plot in two at 6\,\AA\, and
used different Y scalings.  Lower panel show the data/model ratio.}
\label{x}%
\end{figure}

\section{Conclusion: CP\,Pup as a long period intermediate polar}

Our new optical and X-ray data suggest that CP Pup is a magnetic CV, most likely of the intermediate polar (IP) type. In particular, the longer  term  modulation  of  the  radial  velocity  curve 
suggests that CP Pup is not a short orbital period system but a long period CV with orbital period P$\gtrsim$10\,hr. Should this be confirmed, the spectroscopic period of $\sim$1.47\,hr should be interpreted as the WD spin and its instability as an effect induced by the varying geometry of the  emitting  region  (i.e.  the  accretion  curtain  magnetic 
field  lines that are continuously  stretching,  breaking  and reconnecting). The slightly longer photometric period, instead, should be regarded as the beat between the spin and the orbital period. Unfortunately, due to the range of values recorded for both the spectroscopic and photometric period and due to the lack of simultaneous spectroscopic and photometric observations, we could not constrain the putative long orbital period through simple beat frequency calculations. In addition, should CP\,Pup be a long orbital period IP, the observed X-ray modulation (Orio et al. 2009, Balman et al. 1995) would arise from the accreting magnetic pole(s) and match the WD spin period; the NIR light curve (Szkody \& Feinswog 1988) should be interpreted similarly to DQ\,Her (Chanan \& Nelson 1978), and the calculation of the dynamical masses would certainly deliver more reasonable values, possibly consistent with those determined from the X-ray observations. 
However, as the longer period is poorly sampled and highly uncertain, it remains necessary to observe CP\,Pup again, in time resolved spectroscopy and over at least four consecutive nights, thus to securely pin down the long period existence and its exact value.

\acknowledgements  
AB thanks the Space Telescope Science Institute (STScI) for the kind hospitality in June 2012 when this work started. 


\vspace{0.5cm}

{\bf References}

Balman, S., Orio, M., \&  Oegelman, H.,, 1995, ApJ, 449, L47

Barrera, L.H., \& Vogt, N., 1989, RMxA, 19, 99

Bianchini, A., Friedjung, M., \& Sabbadin, F., 1985, IBVS, 2650

Bianchini, A.; Saygac, T.; Orio, M.; Della Valle, M.; Williams, R.E., 2012, A\&A, 539, 94

Chanan, G. A., Nelson, J. E., 1978, ApJ, 226, 963

Cropper, M., 1986, MNRAS, 222, 225

Diaz, M. P., \& Steiner, J. E., 1991, PASP, 103,964

Duerbeck, H. W., Seitter, W. C., \& Duemmler, R., 1987, MNRAS, 260, 149

Mukai, K.; Kinkhabwala, A.; Peterson, J. R.; Kahn, S. M.; Paerels, F., 2003, ApJ, 586, L77

Mukai, K.; Zietsman, E.; Still, M., 2009, ApJ, 707, 652M

O'Donoghue, D., Warner, B., Wargau, W., Grauer, A.D., 1989, MNRAS, 240, 41

Orio, M., Mukai, K., Bianchini, A.,  de Martino, D., \& Howell, S., 2009, ApJ, 690, 1753

Patterson, J., \& Warner, B., 1998, PASP, 110, 1026

Schaefer, B. E.,  \& Collazzi, A. C., 2010, AJ, 139, 1831

Szkody, P., \& Feinswog, L., 1988, ApJ, 334, 422

Warner, B., 1985, MNRAS, 217, 1P

White, J. C., Honeycutt, R.K., \& Horne, K., 1993, ApJ, 412, 278

\end{document}